\documentstyle[12pt]{article}
\begin{document}

\begin{flushright}
THES-TP/92-13\\
December 1993
\end{flushright}
\bigskip

\hrule\hrule

\bigskip

\begin{center}
\Large \bf
 Symmetry algebra of the planar anisotropic quantum harmonic oscillator
 with rational ratio of frequencies
\end{center}

\bigskip\bigskip\bigskip

\centerline{Dennis Bonatsos${}^\dagger$, C.~Daskaloyannis${}^\ast$,
 P.~Kolokotronis${}^\dagger$ and D. Lenis${}^\dagger$}
\bigskip

\centerline{${}^\dagger$ Institute of Nuclear Physics, N.C.S.R.
``Demokritos''}

\centerline{GR-15310 Aghia Paraskevi, Attiki, Greece}

\centerline{${}^\ast$ Department of Physics, Aristotle University of
Thessaloniki}

\centerline{GR-54006 Thessaloniki, Greece }

\begin{abstract}
The symmetry algebra of the two-dimensional quantum harmonic oscillator with
rational ratio of frequencies  is identified as a non-linear extension of
the u(2) algebra. The finite dimensional representation modules  of this
algebra are studied and the energy eigenvalues are determined  using
algebraic methods of general applicability to quantum superintegrable
systems.

\end{abstract}

\bigskip\bigskip\bigskip


\newpage

The two-dimensional anisotropic harmonic oscillator with rational ratio of
 frequencies is a well known example of a classical superintegrable system
 \cite{Hiet1}. It has a third
integral of motion which is given implicitly as a function of the energy
\cite{Contopoulos}, or as a polynomial function of the coordinates and the
momenta \cite{JH}. This oscillator is the natural generalization
of the isotropic harmonic oscillator, both of them widely used in many
branches of physics. The cases with ratio of frequencies 1/2 and 1/3 have
been studied in \cite{Holt} and \cite{FL}, respectively.

The quantum mechanical analogues of the isotropic and anisotropic oscillators
are well known completely solvable models. The symmetry algebra of the
isotropic quantum harmonic oscillator in a space with constant curvature,
 as well as the symmetry algebra of the corresponding Kepler problem,
have been identified in \cite{Higgs,Leemon}. From the algebraic point of
 view, however, while
the many-dimensional analogue of the isotropic quantum oscillator can be
studied using the su(N) or sp(2N,R) algebras, even
for the two-dimensional anisotropic quantum oscillator the symmetry algebra
is still missing.
 Since 1940 Jauch and Hill \cite{JH} have pointed out
that ``the two-dimensional anisotropic oscillator has the same
symmetry as the isotropic oscillator in classical mechanics,
but the quantum mechanical problem presents complications
which leave its symmetry group in doubt''.
The case of ratio of frequencies equal to $1/2$ has been studied in
\cite{Boyer-Wolf}, using the fact that the Hamiltonian is separable in
parabolic
coordinates. The algebra describing the symmetry is a quadratic  Lie algebra
\cite{BDK-PRA}. The importance of the quadratic extensions
of Lie algebras in the study of symmetries was pointed out by
several authors \cite{Zhedanov-AP,Vinet}.
In the case of ratio of frequencies equal to $1/3$ the Hamiltonian is not
separable, but the problem can be solved \cite{BDK-Dubna}, the
symmetry being described by a cubic Lie algebra.
About the general problem of rational ratio of frequencies
Jauch and Hill \cite{JH} noticed that
``the difficulties encountered in this relatively
simple problem throw question on the true interpretation
of classical multiply-periodic motions in quantum mechanics''.

In this letter we identify the symmetry algebra of the
 two-dimensional anisotropic quantum harmonic oscillator, when the
ratio of frequencies is a rational number. The techniques used are related
to the q-deformed oscillator \cite{Biedenharn,Macfarlane,Sun}, introduced
as a tool appropriate for constructing boson realizations of quantum
algebras (quantum groups) \cite{Drinfeld}, which are non-linear
generalizations of the usual Lie algebras. Several mutually equivalent
generalizations of the concept of the deformed oscillator already exist
\cite{generalizations}, their connection to N=2 supersymmetric quantum
mechanics being also discussed \cite{BD-SUSY}. The formalism to be used
here is the one of \cite{D1}. The usefulness of non-linear generalizations
of Lie algebras in the study of integrable systems has been demonstrated
in the case of the Calogero model \cite{Vasiliev,BEM}.

Let us consider the system described by the Hamiltonian:
\begin{equation}
H=\frac{1}{2}\left(
{p_x}^2 + {p_y}^2 + \frac{x^2}{m^2} + \frac{y^2}{n^2} \right),
\label{eq:Hamiltonian}
\end{equation}
where $m$ and $n$ are two natural numbers mutually prime ones, i.e.
their great common divisor is $\gcd (m,n)=1$.

We define the creation and annihilation operators:
\begin{equation}
\begin{array}{ll}
a^\dagger=\frac{x/m - i p_x}{\sqrt{2}}, &
a =\frac{x/m + i p_x}{\sqrt{2}}, \\[0.24in]
b^\dagger=\frac{y/n - i p_y}{\sqrt{2}}, &
b=\frac{y/n + i p_y}{\sqrt{2}}.
\end{array}
\label{eq:operators}
\end{equation}
These operators satisfy the commutation relations:
\begin{equation}
\left[ a,a^\dagger \right] = \frac{1}{m},
\quad
\left[ b,b^\dagger \right] = \frac{1}{n},
\quad
\mbox{other commutators}=0.
\label{eq:commutators}
\end{equation}
Using eqs (\ref{eq:operators}) and (\ref{eq:commutators}) we can prove
by induction that:
$$
\begin{array}{ll}
\left[ a, \left( a^\dagger \right)^p \right] =
 \frac{p}{m} \left( a^\dagger \right)^{p-1} , &
\left[ b, \left( b^\dagger \right)^p \right] =
 \frac{p}{n} \left( b^\dagger \right)^{p-1} ,\\[0.24in]
\left[ a^\dagger, \left( a \right)^p \right] =
- \frac{p}{m} \left( a \right)^{p-1} , &
\left[ b^\dagger, \left( b \right)^p \right] =
- \frac{p}{n} \left( b \right)^{p-1} .
\end{array}
$$
Defining
$$
U=\frac{1}{2} \left\{ a, a^\dagger \right\}, \qquad
W=\frac{1}{2} \left\{ b, b^\dagger \right\},$$
one can easily prove that:
$$\begin{array}{ll}
\left[ U,
\left(a^\dagger \right)^p \right]= \frac{p}{m} \left(a^\dagger \right)^p, &
\left[ W,
\left(b^\dagger \right)^p \right]= \frac{p}{n} \left(b^\dagger \right)^p,
\\[0.24in]
\left[ U,
\left(a \right)^p \right]= - \frac{p}{m} \left(a \right)^p, &
\left[ W,
\left(b \right)^p \right]= - \frac{p}{n} \left(b \right)^p.
\end{array}
$$
Using the above properties we can define the enveloping algebra
generated by the operators:
\begin{equation}
\begin{array}{c}
S_+= \left(a^\dagger\right)^m \left(b\right)^n,\quad
S_-= \left(a\right)^m \left(b^\dagger\right)^n, \\[0.24in]
S_0= \frac{1}{2}\left( U -
W \right), \quad
H=U+W.
\end{array}
\label{eq:generators}
\end{equation}
These genarators satisfy the following relations:
\begin{equation}
\left[ S_0,S_\pm \right]=\pm S_\pm, \quad
\left[H,S_i\right]=0, \quad \mbox{for}\quad i=0,\pm,
\label{eq:SS}
\end{equation}
and
$$
S_+S_- =
\prod\limits_{k=1}^{m}\left( U - \frac{2k-1}{2m} \right)
\prod\limits_{\ell=1}^{n}\left( W + \frac{2\ell-1}{2n} \right),
$$
$$
S_-S_+ =
\prod\limits_{k=1}^{m}\left( U + \frac{2k-1}{2m} \right)
\prod\limits_{\ell=1}^{n}\left( W - \frac{2\ell-1}{2n} \right).
$$
The above relations mean that the harmonic oscillator of eq.
(\ref{eq:Hamiltonian})
is described by the enveloping algebra of the non-linear generalization
of the U(2) algebra formed by the generators $S_0$, $S_+$, $S_-$ and $H$,
satisfying the commutation relations of eq. (\ref{eq:SS}) and
\begin{equation}
\begin{array}{c}
\left[S_-,S_+\right] =
F_{m,n} (H,S_0+1) -  F_{m,n} (H,S_0),\\[0.24 in]
\mbox{where}\quad F_{m,n}(H,S_0)=
\prod\limits_{k=1}^{m}\left( H/2+S_0 - \frac{2k-1}{2m} \right)
\prod\limits_{\ell=1}^{n}\left( H/2-S_0 + \frac{2\ell-1}{2n} \right).
\end{array}
\label{eq:U2}
\end{equation}
This algebra is a non-linear generalization of the U(2) algebra,
of order $m+n-1$. In the case of $m/n=1/1$ this algebra is the
usual U(2) algebra, and the operators $S_0,S_\pm$ satisfy the
commutation relations of the ordinary SU(2) algebra.

The finite dimensional representation modules
 of this algebra can be found
using the deformed oscillator algebra \cite{D1}. The operators:
\begin{equation}
{\cal A}^\dagger= S_+, \quad
{\cal A}= S_-, \quad
{\cal N}= S_0-u, \quad
u=\mbox{ constant},
\label{eq:alge-gen}
\end{equation}
where $u$ is a constant to be determined,
are the generators of a deformed oscillator algebra:
$$
\left[ {\cal N} , {\cal A}^\dagger \right] = {\cal A}^\dagger,
\quad
\left[ {\cal N} , {\cal A} \right] = -{\cal A},
\quad
{\cal A}^\dagger{\cal A} =\Phi( H, {\cal N} ),
\quad
{\cal A}{\cal A}^\dagger =\Phi( H, {\cal N}+1 ).
$$
 The structure function $\Phi$ of this algebra is determined by the function
$F_{m,n}$ in eq. (\ref{eq:U2}):
\begin{equation}
\begin{array}{l}
\Phi( H, {\cal N} )=
F_{m,n} (H,{\cal N} +u ) = \\
= \prod\limits_{k=1}^{m}\left( H/2+{\cal N} +u - \frac{2k-1}{2m} \right)
\prod\limits_{\ell=1}^{n}\left( H/2-{\cal N} -
u + \frac{2\ell-1}{2n} \right).
\end{array}
\label{eq:sf}
\end{equation}
The deformed oscillator corresponding to the structure function of eq.
 (\ref{eq:sf}) has an energy dependent Fock space of dimension $N+1$ if
\begin{equation}
\Phi(E,0)=0, \quad \Phi(E, N+1)=0, \quad
\Phi(E,k)>0, \quad \mbox{for} \quad k=1,2,\ldots,N.
\label{eq:equations}
\end{equation}
The Fock space is defined by:
\begin{equation}
H\vert E, k > =E \vert E, k >, \quad
{\cal N} \vert E, k >= k \vert E, k >,\quad
a\vert E, 0 >=0,
\end{equation}
\begin{equation}
{\cal A}^\dagger \vert E, k> =
\sqrt{\Phi(E,k+1)} \vert E, k+1>,
\quad
{\cal A} \vert E, k> =
\sqrt{\Phi(E,k)} \vert E, k-1>.
\end{equation}
The basis of the Fock space is given by:
$$
\vert E, k >= \frac{1}{\sqrt{[k]!}}
\left({\cal A}^\dagger\right)^k\vert E, 0 >,
\quad k=0,1,\ldots N,
$$
where the ``factorial''  $[k]!$ is defined by the recurrence relation:
$$
 [0]!=1, \quad [k]!=\Phi(E,k)[k-1]! \quad .
$$
Using the Fock basis we can find the matrix representation of the deformed
oscillator and then the matrix representation of the algebra of eqs.
(\ref{eq:SS}), (\ref{eq:U2}).
The solution of eqs (\ref{eq:equations}) implies the following pairs
of permitted values for the energy eigenvalue $E$ and the constant $u$:
\begin{equation}
E=N+\frac{2p-1}{2m}+\frac{2q-1}{2n} ,
\label{eq:E1}
\end{equation}
where $ p=1,2,\ldots,m$, $ q=1,2,\ldots,n$,
and
$$
u=\frac{1}{2}\left( \frac{2p-1}{2m}-\frac{2q-1}{2n} -N \right),
$$
the corresponding structure function being given by:
\begin{equation}
\begin{array}{l}
\Phi(E,x)=\Phi^{N}_{(p,q)}(x)=\\
=\prod\limits_{k=1}^{m}\left( x +
\displaystyle  \frac{2p-1}{2m}- \frac{2k-1}{2m} \right)
\prod\limits_{\ell=1}^{n}\left( N-x+
\displaystyle \frac{2q-1}{2n} +
\frac{2\ell-1}{2n}\right)\\
=\displaystyle\frac{1}{m^m n^n}
\displaystyle\frac{ \Gamma\left(mx+p\right) }{\Gamma\left(mx+p-m\right)}
\displaystyle
\frac{\Gamma\left( (N-x)n + q + n \right)}
{\Gamma\left( (N-x)n + q  \right)}.\end{array}
\label{eq:structure-function}
\end{equation}
In all these equations one has $N=0,1,2,\ldots$, while the dimensionality
of the representation is given by $N+1$. Eq. (\ref{eq:E1})
 means that there are $m\cdot n$ energy eigenvalues corresponding to each
 $N$ value, each eigenvalue having degeneracy $N+1$. (Later we shall see
that the degenerate states corresponding to the same eigenvalue can be
labelled by an ``angular momentum''.)
The energy formula can be corroborated by using the
corresponding Schr\"{o}dinger equation. For the Hamiltonian of eq.
(\ref{eq:Hamiltonian}) the eigenvalues of the Schr\"{o}dinger equation
are given by:
\begin{equation}
E=\frac{1}{m}\left(n_x+\frac{1}{2}\right)+
  \frac{1}{n}\left(n_y+\frac{1}{2}\right),
\label{eq:E2}
\end{equation}
where $n_x=0,1,\ldots$ and $n_y=0,1,\ldots$. Comparing eqs
(\ref{eq:E1}) and (\ref{eq:E2}) one concludes that:
$$N= \left[n_x/m\right]+\left[n_y/n\right],$$
where $[x]$ is the integer part of the number $x$, and
$$
p=\mbox{mod}(n_x,m)+1, \quad q=\mbox{mod}(n_y,n)+1.
$$

The eigenvectors of the Hamiltonian can be parametrized by the
dimensionality of the representation $N$, the numbers $p,q$,
and the number $k=0,1,\ldots,N$:
\begin{equation}
H\left\vert \begin{array}{c} N\\ (p,q) \end{array}, k \right>=
\left(N+\displaystyle
\frac{2p-1}{2m}+\frac{2q-1}{2n}
\right)\left\vert \begin{array}{c} N\\ (p,q) \end{array}, k \right>,
\label{eq:en-rep}
\end{equation}
\begin{equation}
S_0
\left\vert \begin{array}{c} N\\ (p,q) \end{array}, k \right>=
\left(
k+ \displaystyle
\frac{1}{2}
\left( \frac{2p-1}{2m}- \frac{2q-1}{2n} -N \right) \right)
\left\vert \begin{array}{c} N\\ (p,q) \end{array}, k \right>,
\label{eq:s0-rep}
\end{equation}
\begin{equation}
S_+\left\vert \begin{array}{c} N\\ (p,q) \end{array}, k \right>
= \sqrt{ \Phi^N_{(p,q)}(k+1)}
\left\vert \begin{array}{c} N\\ (p,q) \end{array}, k +1\right>,
\label{eq:sp-rep}
\end{equation}
\begin{equation}
S_-\left\vert \begin{array}{c} N\\ (p,q) \end{array}, k \right>
= \sqrt{ \Phi^N_{(p,q)}(k)}
\left\vert \begin{array}{c} N\\ (p,q) \end{array}, k -1\right>.
\label{eq:sm-rep}
\end{equation}

It is worth noticing that the operators $S_0,S_\pm$ do not
correspond to a generalization of the angular momentum,
$S_0$ being the operator corresponding to the Fradkin operator
$S_{xx}-S_{yy}$ \cite{Higgs,Leemon}. The corresponding ``angular
momentum'' is defined by:
\begin{equation}
L=-i\left(S_+-S_-\right).
\label{eq:angular-momentum}
\end{equation}
The ``angular momentum'' operator commutes with the Hamiltonian:
$$ \left[ H,L \right]=0. $$
Let $\vert \ell> $ be the eigenvector of the operator $L$ corresponding
to the eigenvalue $\ell$. The general form of this eigenvector
can be given by:
\begin{equation}
\vert \ell > = \sum\limits_{k=0}^N
\frac{i^k c_k}{\sqrt{[k]!}}
\left\vert \begin{array}{c} N\\ (p,q) \end{array}, k \right>.
\end{equation}

In order to find the eigenvalues of $L$  and the coefficients $c_k$
we use the Lanczos algorithm \cite{Lanczos}, as formulated
 in \cite{Vinet-AP}. From eqs (\ref{eq:sp-rep}) and (\ref{eq:sm-rep}) we find
\begin{equation}
\begin{array}{l}
L\vert \ell > =\ell \vert  \ell >=
\ell\sum\limits_{k=0}^N
\frac{i^k c_k}{\sqrt{[k]!}}
\left\vert \begin{array}{c} N\\ (p,q) \end{array}, k \right>=
\\
=\frac{1}{i}
\sum\limits_{k=0}^{N-1}
\frac{i^k c_k \sqrt{\Phi^N_{(p,q)}(k+1)}}{\sqrt{[k]!}}
\left\vert \begin{array}{c} N\\ (p,q) \end{array}, k+1 \right>-
\frac{1}{i}
\sum\limits_{k=1}^{N}
\frac{i^k c_k \sqrt{\Phi^N_{(p,q)}(k)}}{\sqrt{[k]!}}
\left\vert \begin{array}{c} N\\ (p,q) \end{array}, k-1 \right>
\end{array}
\end{equation} From this equation we find that:
$$c_k=  (-1)^k    2^{-k/2}  H_k (\ell /\sqrt{2} ), $$
where the function $H_k(x)$ is a generalization of the
``Hermite'' polynomials,
satisfying the recurrence relations:
$$
H_{-1}(x)=0, \quad H_0(x)=1,
$$
$$
H_{k+1}(x)= 2 x H_k(x) - 2\Phi^N_{(p,q)}(k) H_{k-1}(x),
$$
and the ``angular momentum'' eigenvalues $\ell$ are the roots of the polynomial
equation:
$$
H_{N+1}(\ell/\sqrt{2}) = 0.
$$
Therefore for a given value of $N$ there are $N+1$ ``angular momentum''
eigenvalues $\ell$, symmetric around zero
 (i.e. if $\ell$ is an ``angular momentum'' eigenvalue,
then $-\ell$ is also an ``angular momentum'' eigenvalue).
In the case of the symmetric harmonic oscillator ($m/n=1/1$)
these eigenvalues are uniformly distributed and differ by 2. In the
general case the ``angular momentum'' eigenvalues are non-uniformly
distributed. Remember that to each value of $N$ correspond $m\cdot n$
energy levels, each with degeneracy $N+1$.

In conclusion,
the two-dimensional anisotropic harmonic oscillator with rational ratio
of frequencies equal to $m/n$
is described dynamically by a non-linear extension of the
u(2) Lie algebra, the order of this algebra being $m+n-1$.
The representation modules of this algebra can be generated
by using the deformed oscillator algebra. The energy eigenvalues are
calculated by the requirement of the existence of finite dimensional
representation modules. The deformed algebra found here to be
the symmetry algebra of the planar anisotropic quantum harmonic oscillator
with rational ratio of frequencies answers the question
posed originally by Jauch and Hill \cite{JH} in 1940.
The next step should be the extension of the proposed algebra to the
three-dimensional and multi-dimensional anisotropic oscillator.

\vfill\eject


\begin{thebibliography}{99}

\bibitem{Hiet1} J. Hietarinta, Phys. Rep. {\bf 147}, 87 (1987).

\bibitem{Contopoulos} G.~Contopoulos, Z. Astrophys. {\bf 49}, 273 (1960);
 Astrophys. J. {\bf 138}, 1297 (1963).

\bibitem{JH} J. M. Jauch and E. L. Hill, Phys. Rev. {\bf 57}, 641 (1940).

\bibitem{Holt} C. R. Holt, J. Math. Phys. {\bf 23}, 1037 (1982).

\bibitem{FL} A. S. Fokas and P. A. Lagerstrom, J. Math. Anal. Appl. {\bf 74},
325 (1980).

\bibitem{Higgs} P. W. Higgs, J. Phys. A {\bf 12}, 309 (1979).

\bibitem{Leemon} H. I. Leemon, J. Phys. A {\bf 12}, 489 (1979).

\bibitem{Boyer-Wolf} C.~P.~Boyer and K.~B.~Wolf, J. Math. Phys. {\bf 16},
2215 (1975).

\bibitem{BDK-PRA} D.~Bonatsos, C.~Daskaloyannis and K.~Kokkotas,
Phys. Rev. A {\bf 48}, R3407 (1993).

\bibitem{Zhedanov-AP} Ya.~I.~Granovskii,   I.~M.~Lutzenko and A.~S.~Zhedanov,
Ann. Phys. (NY) {\bf 217}, 1 (1992).

\bibitem{Vinet} P.  L\'etourneau and L. Vinet, {\it Quadratic Algebras in
Quantum Mechanics}, U. de Montr\'{e}al preprint UdeM-LPN-TH-93-13.

\bibitem{BDK-Dubna} D.~Bonatsos, C.~Daskaloyannis and K.~Kokkotas,
in {\it Symmetry Methods in Physics (Dubna 1993)}, ed. G. Pogosyan, to
appear.

\bibitem{Biedenharn} L. C. Biedenharn, J. Phys. A {\bf 22}, L873 (1989).

\bibitem{Macfarlane} A. J. Macfarlane, J. Phys. A {\bf 22}, 4581 (1989).

\bibitem{Sun} C.~P.~Sun and H.~C.~Fu, J. Phys. A {\bf 22}, L983 (1989).

\bibitem{Drinfeld} V. G. Drinfeld, in {\it Proceedings of the International
Congress of Mathematicians}, edited by A. M. Gleason (American Mathematical
Society, Providence, RI, 1986) p. 798.

\bibitem{generalizations}
J.~Beckers and N.~Debergh, J. Phys. A {\bf 24}, L1277 (1991);
K.~Odaka, T.~Kishi and S.~Kamefuchi, J. Phys. A {\bf 24}, L591 (1991);
A.~Jannussis, G.~Brodimas, D.~Sourlas and V.~Zisis, Lett. Nuovo Cimento
 {\bf 30}, 123 (1981);
G.~Brodimas, A.~Jannussis, D.~Sourlas, V.~Zisis and P.~Poulopoulos,
Lett. Nuovo Cimento {\bf 31}, 177 1981;
M.~Arik, E. Demircan, T. Turgut, L. Ekinci and M. Mungan,
Z. Phys. C {\bf 55}, 89 (1992).

\bibitem{BD-SUSY} D.~Bonatsos and C.~Daskaloyannis, Phys. Lett. B {\bf 307},
 100 (1993).

\bibitem{D1} C. Daskaloyannis, J. Phys. A {\bf 24}, L789 (1991).

\bibitem{Vasiliev} L. Brink, T. H. Hansson and M. A. Vasiliev, Phys. Lett. B
{\bf 286}, 109 (1992).

\bibitem{BEM} T. Brzezi\'nski, I. L. Egusquiza and A. J. Macfarlane,
Phys. Lett. B {\bf 311}, 202 (1993).


\bibitem{Lanczos} C. Lanczos, J. Res. Natl. Bur. Stand. {\bf 45}, 255 (1950).

\bibitem{Vinet-AP} R. Floreanini, J. LeTourneux and L. Vinet, Ann. Phys. (NY)
 {\bf 226}, 331 (1993).

\end{thebibliography}
\end{document}